\documentclass[preprint,prb,aps,floats,amssymb,showpacs,draft,%
floatfix,superscriptaddress]{revtex4} 
\usepackage{psfig}

\def\spose#1{\hbox to 0pt{#1\hss}}
\def\lesssim{\mathrel{\spose{\lower 3pt\hbox{$\mathchar"218$}}
 \raise 2.0pt\hbox{$\mathchar"13C$}}}
\def\gtrsim{\mathrel{\spose{\lower 3pt\hbox{$\mathchar"218$}}
 \raise 2.0pt\hbox{$\mathchar"13E$}}}

\begin{document}
\title{
Zero-temperature behavior of the random-anisotropy model in the 
strong-anisotropy limit }
\author{
Frauke Liers,$^1$ 
Jovanka Lukic,$^2$
Enzo Marinari,$^2$
Andrea Pelissetto,$^2$ and Ettore Vicari$^3$ 
}

\affiliation{Institut f\"ur Informatik,
Pohligstrasse 1, D-50969 K\"oln, Germany \\
$^2$ Dipartimento di Fisica dell'Universit\`a di Roma 
``La Sapienza" and INFN, \\
P.le Aldo Moro 2, I-00185 Roma, Italy. \\
$^3$
Dipartimento di Fisica dell'Universit\`a di Pisa 
and INFN,\\
L.go Pontecorvo 2, I-56100 Pisa, Italy.
}

\date{\today}

\begin{abstract}
We consider the random-anisotropy model on the square and on the cubic
lattice in the strong-anisotropy limit.  We compute exact ground-state
configurations, and we use them to determine the stiffness exponent at
zero temperature; we find $\theta = -0.275(5)$ and $\theta \approx
0.2$ respectively in two and three dimensions.  These results show
that the low-temperature phase of the model is the same as that of
the usual Ising spin-glass model.  We also show that no magnetic order
occurs in two dimensions, since the expectation value of the
magnetization is zero and spatial correlation functions decay
exponentially. In three dimensions our data strongly support the
absence of spontaneous magnetization in the infinite-volume limit.
\end{abstract}


\maketitle


\section{Introduction}
\label{secint}

Amorphous alloys of rare earths, such as Dy, 
and of nonmagnetic transition metals, such as Al, Cu, and Ag, have been 
extensively studied, both theoretically and experimentally.  
They are modeled \cite{HPZ-73} by a Heisenberg model
with random uni-axial single-site anisotropy defined on a simple cubic lattice,
or, in short, by the random-anisotropy model (RAM)
\begin{equation}
{\cal H} =- J \sum_{\langle xy \rangle} \vec{s}_x \cdot \vec{s}_y-
D \sum_x(\vec{u}_x \cdot \vec{s}_x)^2,
\label{HRAM}
\end{equation}
where $\vec{s}_x$ is a three-component spin variable, $\vec{u}_x$ is a unit
vector describing the local (spatially uncorrelated) random anisotropy, and
$D$ is the anisotropy strength. In amorphous alloys the {\em a priori}
distribution of the quenched vectors
$\vec{u}_x$ is usually taken to be isotropic, since, in the absence of
crystalline order, there is no preferred direction.  

Random anisotropy is a relevant perturbation of the pure Heisenberg
model, so that random-anisotropy systems show a critical behavior that
is different from the Heisenberg one. Even though the critical
behavior of the three-dimensional RAM has been investigated at length
in the last thirty years (see Ref.~\onlinecite{DFH-05} for a review),
the phase diagram as a function of $D$ has not yet been determined
conclusively.  The argument of Imry and Ma for $N$-vector systems in
the presence of a random magnetic field \cite{IM-73} has been extended
to the RAM:\cite{PPR-78,JK-80,HCB-87} it forbids the existence of a
low-temperature phase with non-vanishing magnetization for $d<4$. An
analogous conclusion is obtained by considering the
Landau-Ginzburg-Wilson Hamiltonian associated with the
RAM:\cite{DFH-01,CPV-04} no fixed point is found, indicating the
absence of a standard magnetic critical transition.  However, this
does not exclude the possibility of a transition with a
low-temperature phase characterized by magnetic quasi-long-range order
(QLRO), i.e., a phase in which magnetic correlation functions decay
algebraically.\cite{PPR-78} Functional renormalization-group
calculations \cite{Feldman-00,Feldman-01} predict QLRO for small
values of $D$, in agreement with a Landau-Ginzburg calculation of the
equation of state for $D\to 0$.\cite{AP-80} In the large-anisotropy
limit $D\to \infty$ the model becomes an Ising spin glass, in which the
quenched random bond couplings are correlated.  If we write
$\vec{s}_x=\sigma_x \vec{u}_x$ with $\sigma_x=\pm 1$, the RAM reduces
to a particular Ising spin-glass model with Hamiltonian \cite{JK-80}
\begin{equation}
{\cal H} = - \sum_{\langle xy \rangle} j_{xy}\; \sigma_x \; \sigma_y\;,
\qquad j_{xy}\equiv \vec{u}_x \cdot \vec{u}_y\;,
\label{raminfD} 
\end{equation}
which we call strong random-anisotropy model (SRAM) (We set $J=1$
without loss of generality).  Model (\ref{raminfD}) differs from the
usual Ising spin-glass model in the bond distribution.  Here the random variables
$j_{xy}$ on different lattice links are correlated. For instance, one
has $\overline{\prod_\square j_{xy}} = 1/27$, where the product is
over the links belonging to a given plaquette and the average is taken
with respect to the distribution of the vectors $\vec{u}_x$.  An
interesting hypothesis, originally put forward in
Ref.~\onlinecite{CL-77}, is that in this limit the RAM transition is
in the same universality class as that of the Edwards-Anderson Ising
spin-glass model (EAM).\cite{EA,MPV-87,FH-91,KR-03} This conjecture
was confirmed in two dimensions by a renormalization-group calculation
using the large-cell method: the behavior close to the critical point
$T=0$ looks analogous as that of the EAM.\cite{BM-85} In three
dimensions instead the phase diagram has been controversial for a long
time.  While for small values of $D$ numerical simulations
\cite{JK-80,Chakrabarti-87,Fisch-90,Fisch-98,Itakura-03,BCT-05}
confirmed the existence of a finite-temperature transition (though
QLRO was never observed), in the SRAM even the existence of the
transition was in doubt.  \cite{Itakura-03} In
Ref.~\onlinecite{PPV-06} a detailed finite-size scaling study provided
good evidence for the existence of a finite-temperature glassy
transition in the SRAM. Close to the transition, overlap variables,
which are the usual order parameters at a spin-glass transition, are
critical. The corresponding critical exponents are in good agreement
with those obtained for the EAM (see Table 1 in
Ref.~\onlinecite{KKY-06} for a list of recent results) confirming the
conjecture of Ref.~\onlinecite{CL-77}.  The transition in the 
3D SRAM is not a magnetic transition: magnetic variables are not 
critical and on both sides of the transition the system is 
paramagnetic.\cite{PPV-06}

It is interesting to note that Hamiltonian (\ref{raminfD}) is strictly
related to that considered by Hopfield \cite{Hopfield-82-84} in the 
context of neural networks. The main difference lies in the fact 
that in the Hopfield model the components of the vectors $\vec{u}_x$ 
(which are generically $N$ dimensional) are {\em uncorrelated} 
equally distributed random variables, while in the SRAM the different
components are correlated by the constraint $|\vec{u}_x| = 1$. 

The phase diagram of  Hamiltonian (\ref{raminfD}) has been determined 
in the mean-field approximation in Ref.~\onlinecite{DV-80}. 
One finds a critical 
transition followed by a ferromagnetic phase without spin-glass order. 
This result, which is quite general and independent of the nature of 
the distribution of the vectors $\vec{u}_x$, \cite{FZ-85,AGS-85,PV-83}
(apparently, the precise form of the distribution is only relevant for the 
type of magnetic order that sets in as the temperature is lowered below the 
critical point)
is in contrast with the arguments of Ref.~\onlinecite{PPR-78}
and the field-theoretical calculations\cite{DFH-01,CPV-04} and thus does not 
give us any clue on the low-temperature phase.

In this paper we consider the SRAM in two and three dimensions 
and study its behavior at zero
temperature. In particular, we determine the  stiffness exponent 
$\theta$, which is related to the finite-size behavior of the domain-wall 
energy, and several magnetic observables, such as the magnetization,
the susceptibility, and the spin-spin second-moment correlation length. 
For this purpose, by means of an effective 
exact algorithm,\cite{bgjr88,ljrr04} we determine an exact ground state 
for each instance of the randomly chosen vectors $\vec{u}_x$ and 
for different boundary conditions. 

For the stiffness exponent, we find 
$\theta = -0.275(5)$ in two dimensions and $\theta \approx 0.2$ in three
dimensions. These results confirm the conclusions 
of Refs.~\onlinecite{BM-85,PPV-06}, supporting the 
existence of a low-temperature glassy phase in three dimensions analogous 
to that occurring in the EAM and of a two-dimensional
zero-temperature glassy transition in the same universality class as the 
EAM transition with a continuous distribution of the couplings.

As for the magnetic behavior, in two dimensions we can conclude with
confidence that there is no magnetic order: the magnetization vanishes
and magnetic correlation functions decay exponentially with a very
small correlation length, $\xi\approx 2$.  In three dimensions we find
that the magnetization decreases with system size and that the best
fits of the numerical data support the fact that no spontaneous
magnetization occurs in the infinite-volume limit.  This is in
agreement with the results of Ref.~\onlinecite{Fisch-90}, in which a
similar study was presented and no evidence of magnetic criticality
was found.  Since in three dimensions our lattices are relatively
small (even if they are large as compared to state-of-the-art
three-dimensional exact ground-state computations half of the linear
extension of the lattice only amounts to five lattice spacings, which,
together with the need of taking care of finite-size corrections, does
not allow us to distinguish in a clear cut way between a power-law and
an exponential decay) we cannot give a final statement about the issue
of QLRO, though our data are compatible with an exponential decay of
the magnetic correlation functions. As far as we can see, there are no
hints that our model is different from a usual EAM in $3D$.

The paper is organized as follows. In Sec.~\ref{def} we define the quantities
we compute. In Sec.~\ref{mcsim} we present our numerical results: in 
Sec.~\ref{sec-algo} we give some details on the numerical 
methods we use, 
in Sec.~\ref{sec3.A} we compute the stiffness 
exponent, while in Sec.~\ref{sec3.B} we discuss the magnetic 
behavior. Our conclusions are presented in Sec.~\ref{sec4}.

\section{Definitions}
\label{def}

In this work we focus on the computation of the
stiffness exponent $\theta$, of the magnetization of the system 
and of the magnetic correlation functions. 
The exponent $\theta$ is defined in the following way. 
We consider a lattice 
of size
$L^d$ and, for each disorder realization, we compute 
the energies $E_P$ and $E_A$. The energy $E_P$ is the ground-state 
energy for the system with
periodic boundary conditions, whereas the energy $E_A$ is 
the ground-state energy for a system in which anti-periodic boundary conditions
are used in one direction and periodic boundary conditions in the other 
$(d-1)$ directions. As usual, anti-periodic boundary conditions 
are implemented by changing the sign of the bond couplings along 
one lattice $(d-1)$-dimensional boundary. 
More precisely, the model with anti-periodic boundary conditions 
is obtained by considering Hamiltonian (\ref{raminfD}), {\em periodic}
boundary conditions, and couplings 
$j_{x_a x_b} = - \vec{u}_{x_a}\cdot \vec{u}_{x_b}$
when $x_a = (1,n_2,\ldots, n_d)$ and 
$x_b = (L,n_2,\ldots, n_d)$.\cite{footnote-AP} Then, we define 
\begin{equation}
E_m \equiv \overline {E_P - E_A} \qquad\qquad 
\Delta E = \overline {|E_P - E_A - E_m|},
\label{defDeltaE}
\end{equation}
where the over-line indicates the average over the distribution of the
vectors $\vec{u}_x$. Note that in the definition we have subtracted
the non-zero average $E_m$. Only with this subtraction does $\Delta E$
provide a measure of the width of the domain-wall distribution.  The
presence of $E_m$ in the definition deserves some comments.  In the
usual EAM, $E_m = 0$. Indeed, the bond distribution is invariant under
the change of sign of any number of couplings, so that $E_A$ and $E_P$
have the same distribution, which implies $\overline{E}_A =
\overline{E}_P$ and therefore $E_m = 0$.  Thus, this subtraction is
not needed in the EAM definition of $\Delta E$.  

In the SRAM, instead, this symmetry does not hold.  To understand why
we first notice that the products of couplings over closed loops that
do not wrap around the lattice (trivial loops) is the same when using
periodic or antiperiodic boundary conditions, since in any such loop
one always gets an even number of sign changes.  Consider now the
product $P(n_2,\ldots,n_d) = j_{x_1x_2} j_{x_2x_3}\ldots j_{x_Lx_1}$,
where $x_k = (k,n_2,\ldots,n_d)$, i.e. the product of the bond
couplings along one line (which is frequently known as Polyakov line)
that wraps around the lattice in the
direction
where  antiperiodic boundary conditions have been imposed.
Averaging over the $\{u_x\}$ distribution we obtain
\[
\overline{P(n_2,\ldots,n_d)} = 3^{1-L}.
\]
When we consider antiperiodic boundary conditions we change 
the sign of one of the links belonging to the Polyakov line, 
and thus in this case the 
average of $P(n_2,\ldots,n_d)$ is $-3^{1-L}$. This indicates 
that the probability distribution of the bond couplings 
for periodic and antiperiodic boundary conditions is different.
Thus, we have $\overline{E}_A \not= \overline{E}_P$, which 
implies $E_m\not=  0$.
Because of that when
subtracting $E_m$, $\Delta E$ provides a measure of the width of the
domain-wall distribution.

For $L\to \infty$, $\Delta E$ behaves as 
\begin{equation}
\Delta E \sim L^\theta,
\end{equation}
which defines the exponent $\theta$.

We also consider magnetic correlations. They are defined in terms 
of the variables $\vec{s}_x = \sigma_x \vec{u}_x$. In particular,
we consider the average absolute value of the
magnetization per site
\begin{equation}
m = {1\over V} \overline{ \left \langle \left|
   \sum_{x} \vec{s}_x \right| 
   \right \rangle },
\end{equation}
the spin-spin correlation function 
\begin{equation}
G(x) \equiv \overline{ \langle \vec{s}_0\cdot \vec{s}_x\rangle } - m^2 = 
      \overline{ \vec{u}_0\cdot\vec{u}_x \langle \sigma_0\sigma_x\rangle } -
      m^2, 
\end{equation}
its Fourier transform $\widetilde{G}(p)$, 
the corresponding susceptibility $\chi$, and the second-moment
correlation length $\xi$:
\begin{eqnarray}
&& \chi \equiv  \sum_{x} G(x) = \widetilde{G}(0), \qquad\qquad
\\
&& \xi^2 \equiv  {1\over 4 \sin^2 (p_{\rm min}/2)} 
{\chi - F\over F}, \qquad\qquad 
F \equiv  \widetilde{G}(p) = \sum_x G(x) \cos {2 \pi x_1\over L},
\label{xidefffxy}
\end{eqnarray}
where $p = (p_{\rm min},0,0)$, and $p_{\rm min} \equiv 2 \pi/L$.

\section{Results}
\label{mcsim}

\begin{table}
  \begin{tabular}{cccc}\hline\hline
     \multicolumn{2}{c}{$d = 2$} &  \multicolumn{2}{c}{$d = 3$} \\
     $L$ & $N_0(L)$ & $L$ & $N_0(L)$ \\ \hline
     \multicolumn{1}{c}{\hphantom{???}$L\leq 60$} &  $10000$  &
     \multicolumn{1}{c}{\hphantom{???}$L\leq 6$} &  $20000$
         \\
     $70$      & $5000$     & $7 $     &$14000$ \\
     $80$      & $4000$     & $8 $     &$18000$  \\
     $90$      & $4000$     & $9 $     &$13860$ \\
     $100$     & $3600$     & $10$     &$4479$ \\
     $110$     & $1600$ \\
     $120$     & $1000$ \\ 
  \hline\hline
  \end{tabular}
  \caption{Number $N_0(L)$ 
     of computed ground states for two- ($d=2$) and three-dimensional
     ($d=3$) lattices.}
  \label{tab:numbersamples}
\end{table}

\subsection{The algorithm} \label{sec-algo}

At zero temperature the determination of the thermal averages reduces
to the evaluation of the observables in the ground-state
configuration.  We determine an exact ground state by computing a
maximum cut in the interaction graph.\cite{ba82} This is a prominent
problem in combinatorial optimization, which, for general graphs, is
NP-hard.  
However, it can be solved in polynomial time when restricted
to two-dimensional lattices 
with either free boundaries or periodic boundary conditions
where the coupling sizes $j_{xy}$ (assumed integer) are bounded by a
polynomial in the size of the input.  For the case of continuous
couplings that we consider here the complexity status is not known.  

For three-dimensional instances, the problem is NP-hard independent of
the boundary conditions. For the SRAM model considered here, we use a
branch-and-cut approach that is especially designed for solving
NP-hard instances. \cite{bgjr88,ljrr04}

To compute an exact ground state, we consider the lattice as a graph $G=(V,E)$,
in which the nodes $V$ are the lattice sites and the edges $E$ are the lattice 
links that correspond to a non-vanishing coupling (in our case, only nearest
neighbors are connected). To each edge we associate a {\em cost}: 
the cost $c_{u,v}$ of an edge $(u,v)\in E$ is the negative coupling strength 
$-j_{uv}$.
Given a partition of the nodes into 
two sets $W$ and $V\setminus W$, we associated to it a cut in $G$,
which is an edge set that contains 
all edges $e=(u,v)$ such that 
$u\in W$ and $v\in V\setminus W$. To each cut we associate a {\em cut value}, 
which is the sum of the costs of the cut edges. 
It is not hard to see that a ground state
can be obtained as follows. One first determines a maximum cut in $G$, that 
is a cut which has a maximal value among all possible cuts.
Then, a ground-state
spin configuration is obtained by assigning one orientation to the spins
that belong to one of the node partitions and the opposite orientation to the
others.

To determine a maximum cut, we use a branch-and-cut algorithm from
combinatorial optimization. By studying the geometric structure of
the problem, we can derive upper bounds for the maximum-cut value. 
A lower bound is given by the value of any cut. 
During the run of the algorithm, we
iteratively improve upper and lower bounds on the
problem's solution value. It can happen that one cannot improve these
bounds any further. In this case we split up the problem into easier
sub-problems, which we solve recursively by improving their
corresponding upper and lower bounds. We continue the process of
tightening the bounds and splitting up the problem into easier sub-problems 
until upper and lower bounds coincide.
This provides 
an optimal solution and a ground state of the system.
Note that in the presence of degeneracies
the algorithm finds only one of the ground states. However, since 
in our case the bond couplings are real numbers, we do not expect degeneracies
and thus the algorithm finds the unique ground state.

This exact algorithm allows us to compute the ground state on 
square lattices $L^2$, $5\le L \le 120$ and 
on cubic lattices $L^3$, $3\le L \le 10$ 
within reasonable time.
For a two-dimensional lattice with $L\leq 80$ and periodic boundary conditions,
one ground-state computation takes less than two minutes on average on
a SUN Opteron (2.2 GHz) machine; for $120^2$ lattices the 
computation takes 28 minutes.
Solving the problem for three-dimensional lattices is more
difficult, especially for periodic boundary conditions as we use
here. One ground-state computation
takes less than 20 seconds for $L\le 8$, whereas the average
CPU time is $8$ minutes for $L=10$. We
report the number of computed samples in Table~\ref{tab:numbersamples}.

\subsection{Stiffness exponent} \label{sec3.A}

\begin{figure*}[tb]
\centerline{\psfig{width=12truecm,angle=-90,file=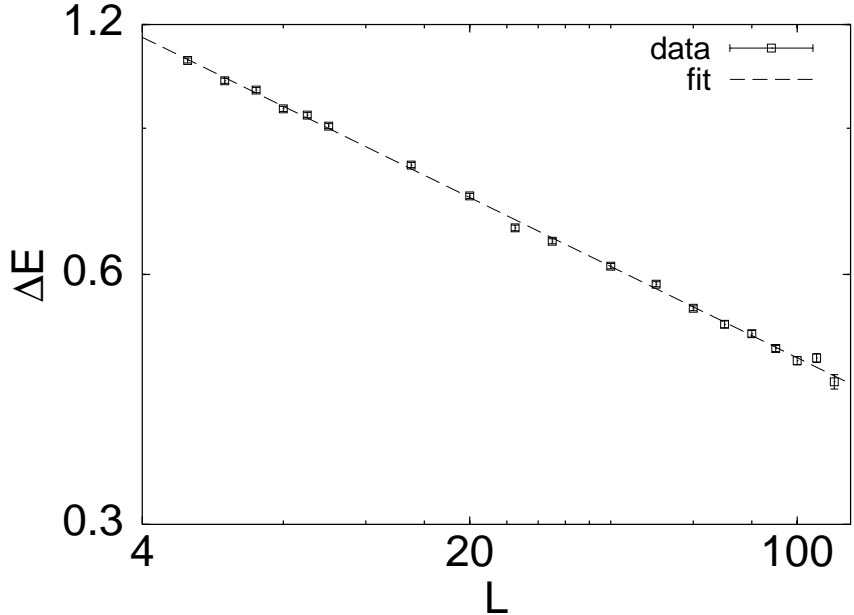}}
\vspace{2mm}
\caption{Estimates of $\Delta E$ in two dimensions. We also report
the curve $a L^\theta$, $a = 1.699$, $\theta = -0.276$,
 obtained by fitting all data.
}
\label{figDeltaE-2D}
\end{figure*}

\begin{figure*}[tb]
\centerline{\psfig{width=12truecm,angle=-90,file=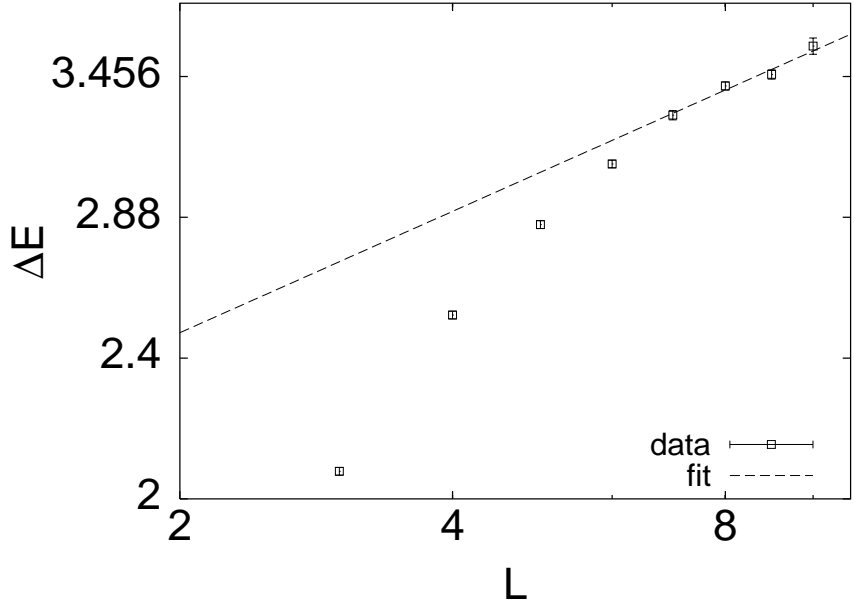}}
\vspace{2mm}
\caption{Estimates of $\Delta E$ in three dimensions. We also report
the curve $a L^\theta$ obtained by fitting 
the last four data points 
($L_{\rm min} = 7$), $a = 2.12$, $\theta = 0.227$.
}
\label{figDeltaE-3D}
\end{figure*}

\begin{table}[tbp]
\caption{Estimates of $\theta$ in two dimensions. 
We also report the square of the residuals ($\chi^2$) and its value 
divided by the number of degrees of freedom (DOF). 
}
\label{tabletheta2D}
\begin{center}
\begin{tabular}{cccc}
\hline\hline
\multicolumn{1}{c}{$L_{\rm min}$}&
\multicolumn{1}{c}{$\theta$} &
\multicolumn{1}{c}{$\chi^2$} &
\multicolumn{1}{c}{$\chi^2$/DOF} \\
\hline 
5    &  $-0.276(2)$ &  29 & 1.7\\
10   &  $-0.278(3)$ &  24 & 2.0\\
20   &  $-0.271(4)$ &  19 & 1.9\\
30   &  $-0.271(7)$ &  11 & 1.4\\
40   &  $-0.279(9)$ &   9 & 1.3\\
\hline\hline
\end{tabular}
\end{center}
\end{table}

We have measured the stiffness exponent in two and in three dimensions.
Estimates of $\Delta E$ on square lattices $L^2$, $5\le L \le 120$
are reported in Fig. \ref{figDeltaE-2D} versus $L$. On a logarithmic scale
the data fall
on a straight line quite precisely. If we fit $\Delta E$ to 
\begin{equation}
 \ln \Delta E = a + \theta \ln L,
\label{DeltaE}
\end{equation}
including only data with $L\ge L_{\rm min}$, we obtain the results 
reported in Table~\ref{tabletheta2D}.
No significant scaling corrections are present
and the estimate of $\theta$ is constant within error bars. We take 
as our final estimate 
\begin{equation}
\theta = - 0.275(5),
\label{theta-2D}
\end{equation}
which includes all results.
Estimate (\ref{theta-2D}) should be compared with 
those obtained for the EAM with continuous energy distributions
(if energies are quantized the stiffness exponent vanishes,
see the discussion in Ref.~\onlinecite{AMMP-03}):
$\theta = -0.281(2)$ (Ref.~\onlinecite{Rieger-etal-96-97}),
$\theta = -0.282(2)$ (Ref.~\onlinecite{HY-01}),
$\theta = -0.282(3)$ (Ref.~\onlinecite{CBM-02}).
Our result is consistent, indicating that the $T=0$ transition
in the SRAM belongs to the same universality class as that 
of the EAM, as found in Ref.~\onlinecite{BM-85}.

\begin{table}[tbp]
\caption{Estimates of $\Delta E$, $m$, $\chi$, and $\xi^2$
in three dimensions. 
}
\label{tablemagnetic3D}
\begin{center}
\begin{tabular}{ccccc}
\hline\hline
\multicolumn{1}{c}{$L$}&
\multicolumn{1}{c}{$\Delta E$} &
\multicolumn{1}{c}{$m$} &
\multicolumn{1}{c}{$\chi$} &
\multicolumn{1}{c}{$\xi^2$} \\
\hline 
3 & 2.073(10) & 0.5601(6) &  & \\
4 & 2.538(13) & 0.4985(5) &  & \\
5 & 2.853(15) & 0.4468(5) & 0.3446(9) &  $-1.716(3)$ \\
6 & 3.086(15) & 0.4022(6) & 0.5151(11) &  $-3.320(7)$ \\
7 & 3.287(20) & 0.3638(6) & 0.5213(12) &  $-12.15(7)$ \\
8 & 3.414(18) & 0.3292(7) & 0.6277(12) &  $-35.13(3)$ \\
9 & 3.465(21) & 0.2986(10)& 0.6276(12) &  $13.82(8)$ \\
10& 3.595(38) & 0.2710(16)& 0.6915(24) &  $12.86(13)$ \\
\hline\hline
\end{tabular}
\end{center}
\end{table}

\begin{table}[tbp]
\caption{Estimates of $\theta$ in three dimensions. 
We also report the square of the residuals ($\chi^2$) and its value 
divided by the number of degrees of freedom (DOF). 
}
\label{tabletheta3D}
\begin{center}
\begin{tabular}{cccc}
\hline\hline
\multicolumn{1}{c}{$L_{\rm min}$}&
\multicolumn{1}{c}{$\theta$} &
\multicolumn{1}{c}{$\chi^2$} &
\multicolumn{1}{c}{$\chi^2$/DOF} \\
\hline 
3   &  0.465(5) &  280 & 46.7\\
4   &  0.390(7) &   70 & 13.9\\
5   &  0.338(11) &  26 & 6.4\\
6   &  0.294(16) &  12 & 3.9\\
7   &  0.227(28) &  2.7 &1.3\\
8   &  0.197(47) &  2.1 &2.1\\
\hline\hline
\end{tabular}
\end{center}
\end{table}

We have repeated the analysis in three dimensions. Estimates of 
$\Delta E$ on a cubic lattice $L^3$, $3\le L \le 10$, are reported in
Table~\ref{tablemagnetic3D} and plotted in Fig.~\ref{figDeltaE-3D}. 
The energy difference $\Delta E$ increases with $L$, indicating that
$\theta > 0$. This in turn implies the existence of a low-temperature 
glassy phase and of a finite-temperature glassy transition,
confirming the results of Ref.~\onlinecite{PPV-06}. In order to determine 
$\theta$ we performed fits of the form (\ref{DeltaE}). 
The results, corresponding to
different values of $L_{\rm min}$, are reported in Table~\ref{tabletheta3D}. 
In this case there are significant scaling 
corrections: the $\chi^2$ is large for small values of $L_{\rm min}$ 
and a significant downward trend is visible in the estimates of $\theta$. 
A reasonable $\chi^2$ is obtained for $L_{\rm min} \ge 7$, corresponding to
$\theta\approx 0.2$. It is difficult to set a reliable error bar on this 
value. Nonetheless, let us note that this estimate is close 
to all results obtained for the EAM.
A determination of 
$\theta$ on cubic lattices as done here gives 
$\theta = 0.19(2)$ (Ref.~\onlinecite{Hartmann-99}) and 
$\theta \approx 0.19$ (Ref.~\onlinecite{CBM-02}), while the aspect-ratio 
scaling  method gives a slightly different result \cite{CBM-02}
$\theta \approx 0.27$. Given the uncertainties of the EAM results
and the relatively small lattice sizes considered in our investigation,
we can certainly conclude that our estimate of $\theta$
is fully compatible with the EAM one, confirming 
the findings of Ref.~\onlinecite{PPV-06}.


\subsection{Magnetic behavior} \label{sec3.B}

\begin{figure*}[tb]
\centerline{\psfig{width=12truecm,angle=-90,file=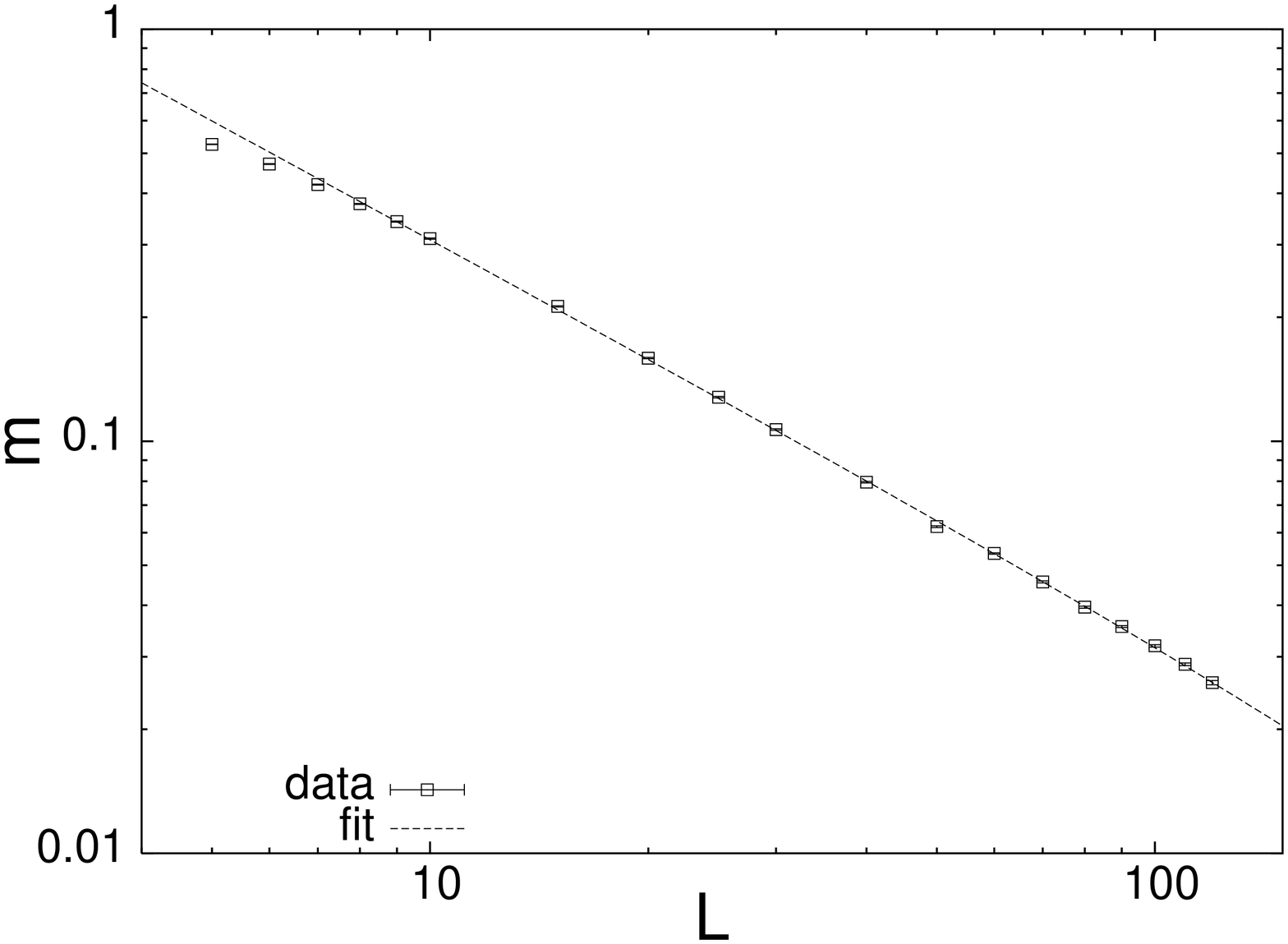}}
\vspace{2mm}
\caption{Estimates of the magnetization $m$ in two dimensions. We also report
the curve obtained by fitting all data.
}
\label{figm-2D}
\end{figure*}

\begin{figure*}[tb]
\begin{tabular}{cc}
\psfig{width=7truecm,angle=-90,file=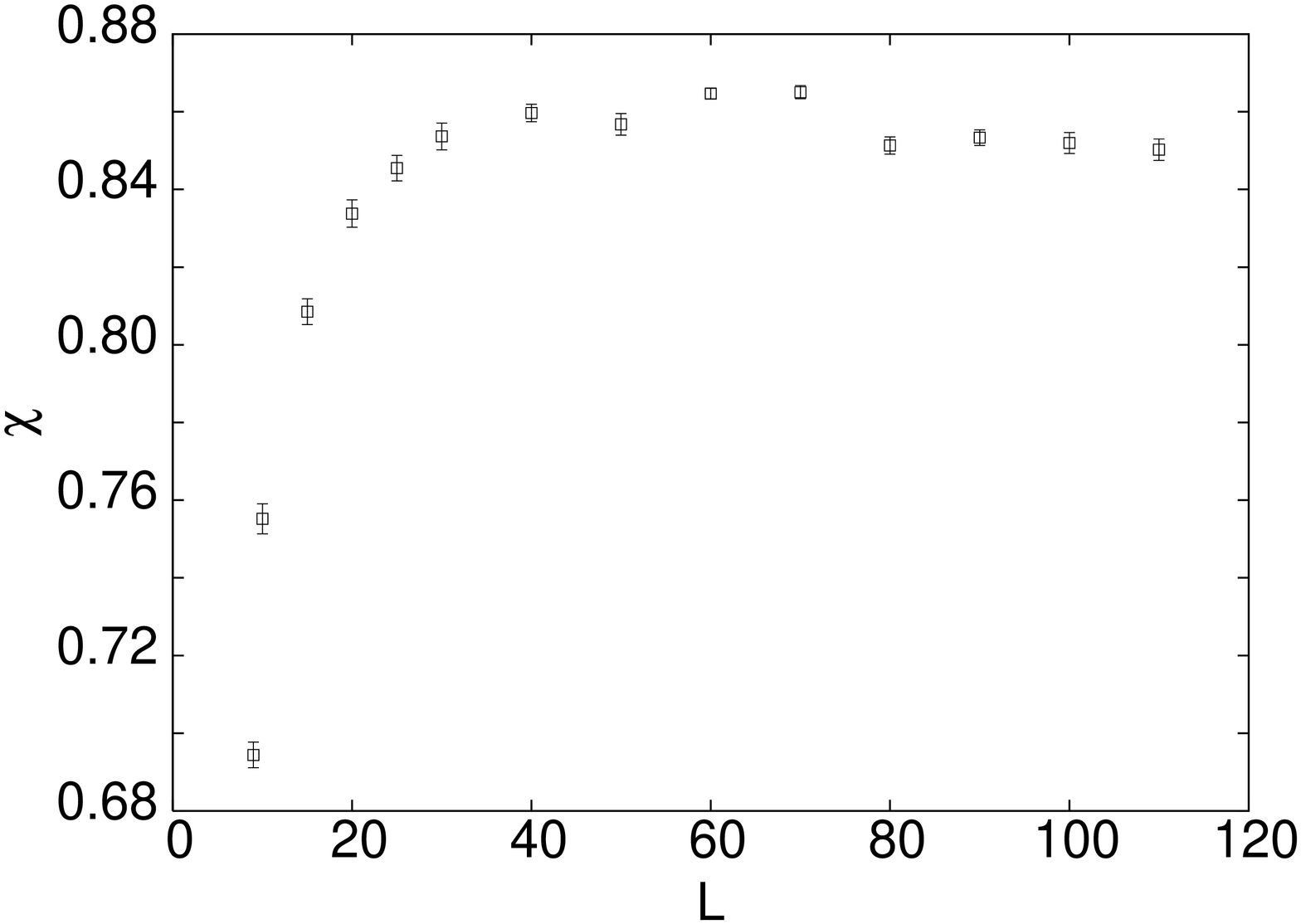} &
\psfig{width=7truecm,angle=-90,file=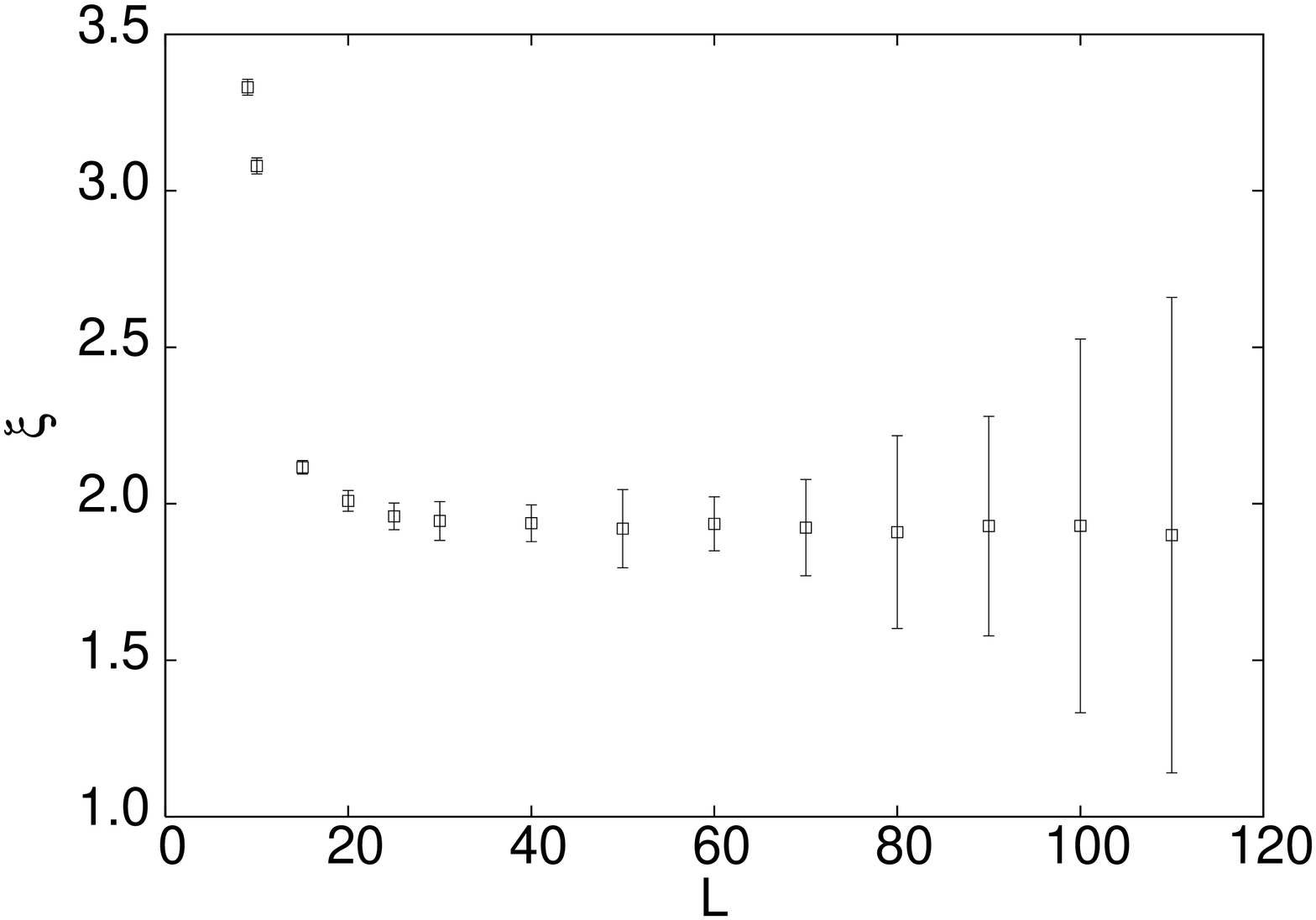} \\
\end{tabular}
\vspace{2mm}
\caption{Estimates of the susceptibility and of the correlation length
in two dimensions. 
}
\label{figchi-xi-2D}
\end{figure*}

\begin{figure*}[tb]
\centerline{\psfig{width=10truecm,angle=-90,file=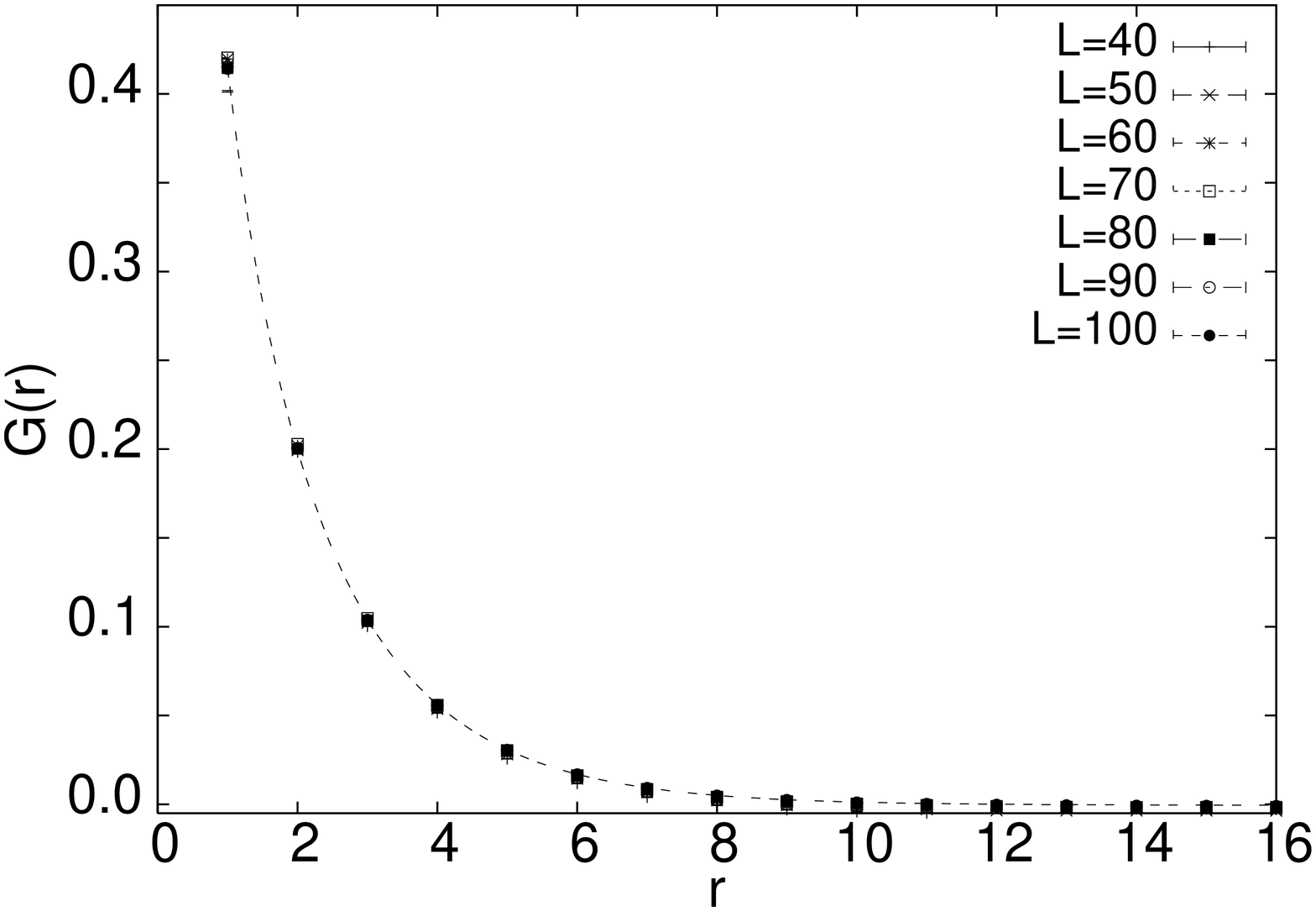}}
\vspace{2mm}
\caption{Connected magnetic correlation function $G(r)$
in two dimensions. 
}
\label{figGc-2D}
\end{figure*}

Once it has been established that the SRAM has a glassy ground state,
it is of interest to check whether at $T=0$ glassy behavior and 
some kind of magnetic order coexist. 

In Fig.~\ref{figm-2D} we show the average magnetization per site $m$ versus 
$L$ in two dimensions. The magnetization decreases as expected. 
Moreover a fit of $\ln m$ to $a + \rho \ln L$ gives $\rho \approx -1$. 
More precisely, we obtain $\rho =-0.9405(8)$,  $-0.9946(17)$,
$-1.003(3)$  for $L_{\rm min} = 5,10,20$,
respectively. These results are perfectly consistent with a behavior of the 
form $m\sim V^{-1/2}$, where $V$ is the volume, which is the 
expected behavior if the system is paramagnetic. As a check we also 
computed $\chi$ and $\xi$, which are reported in Fig.~\ref{figchi-xi-2D}. 
They become constant as $L\to \infty$ indicating the absence of 
magnetic order. Moreover, $\chi$ converges to a constant with 
$1/V$ corrections, as expected: indeed, a fit of $\chi$ to $a + b/L^\delta$
gives $a = 0.8617(8), 0.8607(9)$ and 
$\delta = 2.02(2), 2.18(13)$ for $L_{\rm min} = 5$, 10. Analogously,
$\xi^2$ converges to $\xi = 1.90(5)$: magnetic correlations extend only
over two lattice spacings.
Finally, in Fig.~\ref{figGc-2D} we report
$G(r)$ for several values of $L$. 
No $L$ dependence can be observed, so that our data provide the 
infinite-volume spin-spin correlation function.
In two dimensions and in infinite volume we expect 
\begin{equation}
G(r) \approx {A\over \sqrt{r}} e^{-r/\xi_e}
\end{equation}
for $r\to \infty$, where $\xi_e$ is a second definition of 
correlation length. Fitting the data in the range $r\in [a,b]$, 
$a\approx 3$-6, $b\approx 13$-15, for $L\ge 60$, we always obtain 
$\xi_e \approx 2$, which is, as expected, close to the estimate of 
the second-moment correlation length considered before.
Clearly, for $T=0$ the system is not magnetized nor is there QLRO.

Let us now consider the three-dimensional case. The mean values of 
the magnetization, $\chi$, and $\xi^2$ are reported in Table
\ref{tablemagnetic3D}. The magnetization decreases, as already observed in 
Ref.~\onlinecite{Fisch-90}, thus supporting the claim that no spontaneous 
magnetization occurs. Fits that lead to a non-magnetized 
infinite-volume limit are always preferred to best fits that imply a
spontaneous magnetization: if we fit the data to the form
$m + a L^{-x}$,
fixing $m$ to a given value (we have tried for example $m=0.05$, $0.1$
and $0.15$), the reduced $\chi^2$ decreases with decreasing 
(fixed) values of $m$.
Also a fit of the correlation functions to an exponential decay 
has a better $\chi^2$
than a fit to a pure power law (always considering fits with
the same number of parameters).

\begin{figure*}[tb]
\centerline{\psfig{width=10truecm,angle=-90,file=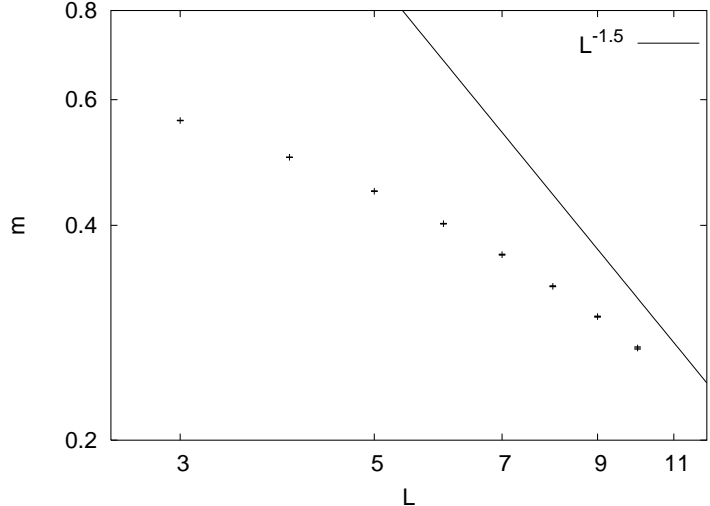}}
\vspace{2mm}
\caption{Log-log plot of the magnetization $m$ in three dimensions 
as a function of $L$.}
\label{fig1-m-3D}
\end{figure*}

\begin{figure*}[tb]
\centerline{\psfig{width=10truecm,angle=-90,file=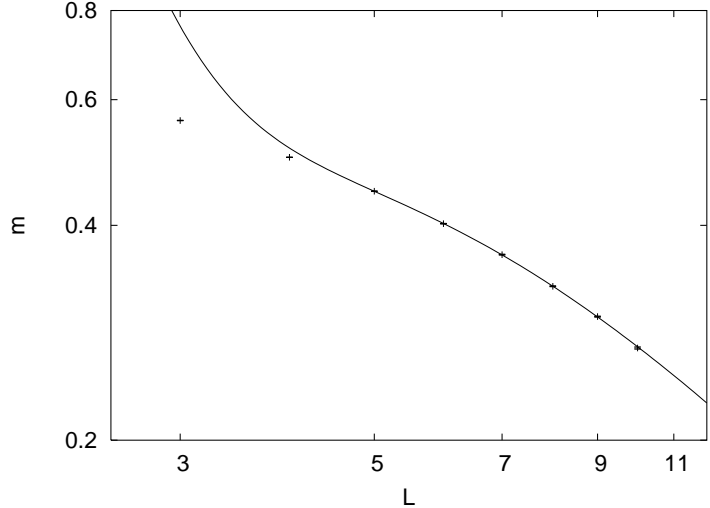}}
\vspace{2mm}
\caption{Three-dimensional average magnetization 
versus $L$ on a log-log scale. The continuous line is the best fit to 
(\ref{fit-3D}), which accounts for finite-size corrections (only data
with $L \geq 5$ have been considered in the fit).}
\label{fig2-m-3D}
\end{figure*}

The presence of large finite-size corrections does not allow us to 
verify the expected asymptotic behavior
$m \sim V^{-1/2} \sim L^{-3/2}$.
However, as we show in Fig.~\ref{fig1-m-3D}, 
the data show a clear trend compatible with 
this behavior.
To make a more quantitative comparison we have checked that the deviations
can be interpreted as scaling corrections.
For this purpose we fit the data with $L\ge 5$ to 
\begin{equation}
\frac{A}{L^{1.5}}\left (1+\frac{B}{L}+\frac{C}{L^2}\right)\;,
\label{fit-3D}
\end{equation}
including two analytic corrections. If the system is paramagnetic non-analytic 
exponents are not expected and thus Eq.~(\ref{fit-3D}) represents the 
expected asymptotic form. The fit---the resulting curve is  shown in 
Fig.~\ref{fig2-m-3D}---is quite good and provides very reasonable 
values for the fit parameters: $A\simeq 15$, $B\simeq -5$, and $C\simeq 8$.

In three dimensions we cannot draw any final
conclusion on the question of QLRO 
from the data of $\chi$ and $\xi$, since currently treatable lattice sizes
are too small
to allow a clear-cut selection of a given functional behavior.
We present here a few comments.
First, the values we find 
for $\chi$ are quite small, of the same order of those 
occurring in two dimensions, where we know with confidence 
that there is no magnetic critical behavior. Second, note that for $L\le 8$, 
$\xi^2$ is negative. This happens because $F$ [see definition 
(\ref{xidefffxy})]
is small and negative ($F\approx -0.03$ for $L = 8$), indicating that 
there is no magnetic order, even on a scale of one lattice spacing.
For $L = 9,10$ we find 
$\xi \approx 3.7$ (the approximate equality of the two values 
is probably an effect of 
even-odd oscillations, which are typical of systems with anti-ferromagnetic 
couplings, and should not be taken as an indication that $\xi$ is already
close to its infinite-volume value $\xi_\infty$). Since 
infinite-volume results can only be obtained if $L\gtrsim c\;\xi$, 
$c\gtrsim 4$-5, we expect
that lattices with at least $L = 20$ are needed in order to 
give a definite assessment about
the question of magnetic QLRO in three dimensions. 

\section{Conclusions} \label{sec4}

In this paper we have investigated the behavior of the SRAM at $T=0$
in two and three dimensions. Our main results are the following:
\begin{itemize}
\item[(i)]  We determine the stiffness exponent, obtaining 
$\theta \approx 0.2$ in three dimensions and $\theta = -0.275(5)$ in 
two dimensions. These results show that the low-temperature behavior of the 
SRAM is the same as that of the EAM, confirming the 
conclusions of Refs.~\onlinecite{PPV-06,BM-85}. In particular, the 
correlation among the bond couplings is irrelevant.
\item[(ii)] We investigate the question of the magnetic order. 
In two dimensions we find no evidence of critical behavior: magnetic 
correlations die out after a few lattice spacings. In three dimensions
we exclude the presence of spontaneous magnetization, in agreement
with Ref.~\onlinecite{Fisch-90}. The question of QLRO is still open;
the limited linear size only allows us to claim that the decay
of correlation functions is compatible with an exponential
decay.
Note that 
if QLRO would hold 
at $T=0$, a second transition should occur, at temperatures 
below the temperature $T_g$ of the glassy transition found in 
Ref.~\onlinecite{PPV-06}. Indeed, the numerical data 
of Ref.~\onlinecite{PPV-06} indicate paramagnetic
behavior all around $T_g$.
\end{itemize}
There are several generalizations of the SRAM that can be investigated 
with the method we use here. For instance, we could consider 
$N$-dimensional vectors $u_x$ with $N\not=3$ 
or different distributions of the vectors $u_x$. 
In the first case, we can give precise predictions.
The correlation of the bond variables around a lattice
plaquette becomes $\overline{\prod_\square j_{xy}} = 1/N^3$, which implies
that bond correlations vanish for $N\to \infty$. Thus, for $N = \infty$, the
SRAM is just an EAM with a different continuous bond distribution.  In this
limit, therefore, the two models belong to the same universality class.  
Our results for $N=3$ imply that the same holds for any $N\ge 3$.
For $N=1$ it is enough to redefine $\sigma_i \to u_i \sigma_i$ to re-obtain the
standard ferromagnetic Ising model. The behavior for $N=2$ is  
not predicted by our results, 
since, for $N = 2$, the model is less frustrated than that with $N=3$ 
studied here. In three dimensions,
numerical studies \cite{FH-90,Fisch-90,Fisch-91,Fisch-95} 
provide some evidence that the $N=2$ SRAM  has a magnetic transition with 
a diverging magnetic susceptibility. The nature of the low-temperature
phase is however still controversial.

Little is known for generic distributions of the vector $u_x$.  The
arguments of Refs.~\onlinecite{PPR-78,JK-80} do not necessarily apply
to this case.  Indeed, they either assume that correlation functions
have a Goldstone-like singularity or that the relevant magnetic modes
are spin waves. Both assumptions may not hold for generic
distributions, since the $O(N)$ symmetry is broken even after
averaging over disorder.  The only available results are those of
Ref.~\onlinecite{CPV-04} that considers generic cubic-symmetric
distributions in three dimensions.  They generically exclude the presence of a
ferromagnetic transition belonging to the random-exchange universality
class (there are some exceptions, but they appear to be of limited
practical interest \cite{footnote1}). Different types of magnetic
transitions are however still possible, and in this case nothing is
known on a possible glassy transition and on the presence of QLRO.

\section*{Acknowledgments}

We thank Silvio Franz for an interesting conversation.
The computations were performed on the Cliot cluster of the Regional
Computing Center and on the scale cluster of E. Speckenmeyer's group,
both at the University of K\"oln, Germany. 
F.L. has been supported by the German Science
Foundation (DFG) in the projects Ju 204/9 and Li 1675/1 and by the Marie
Curie RTN ADONET 504438 funded by the EU.


\begin{thebibliography}{999}
\bibitem{footnote-AP}
We could impose antiperiodic boundary conditions directly on model
(\ref{HRAM}), i.e., set $\vec{s}_{L+1,\ldots} = - \vec{s}_{1,\ldots}$ and 
$\vec{u}_{L+1,\ldots} = - \vec{u}_{1,\ldots}$. This amounts to consider 
model (\ref{HRAM}) with interaction $+J \vec{s}_{x_a} \cdot \vec{s}_{x_b}$,
when $x_a = (1,\ldots,n_d)$ and $x_b = (L,\ldots,n_d)$. Under the mapping
$\vec{s}_x = \vec{u}_x \sigma_x$, we reobtain the same Hamiltonian
discussed in the text.
\end{thebibliography}
\end{document}